\documentclass[11pt]{article} 
\usepackage{graphicx,psfig,floatflt,amssymb,epsf,rotate} 
\def\gtap{\raisebox{-.4ex}{\rlap{$\sim$}} \raisebox{.4ex}{$>$}} 
 
\def\fprime{\hspace*{-2pt}\raisebox{0.65ex}{\tiny$'$}\hspace{2pt}}
\begin{document} 
\begin{flushright} 
SINP/TNP/09-17, ~ CU-Physics-05/2009, ~ CP3-09-13, ~
HRI-P-09-03-002, ~ RECAPP-HRI-2009-011
\end{flushright} 
 
\vskip 30pt 
 
\begin{center} 
{\Large \bf Exploring the Universal Extra Dimension at the LHC} \\
\vspace*{1cm} 
\renewcommand{\thefootnote}{\fnsymbol{footnote}} 
{ {\sf Gautam Bhattacharyya${}^1$}, {\sf Anindya Datta${}^{2}$}, 
{\sf Swarup Kumar Majee${}^{3}$}, {\sf Amitava Raychaudhuri${}^{4,2}$} 
} \\ 
\vspace{10pt} 
{\small ${}^{1)}$ {\em Saha Institute of Nuclear Physics, 
1/AF Bidhan Nagar, Kolkata 700064, India} \\
   ${}^{2)}$ {\em Department of Physics, University of Calcutta, 
92 A.P.C. Road, Kolkata 700009, India}  \\
   ${}^{3)}$ {\em Center for Particle Physics and Phenomenology (CP3), 
Universite\fprime Catholique de Louvain, \\ Chemin du Cyclotron 2, 
B-1348 Louvain-la-Neuve, Belgium}\\
   ${}^{4)}$ {\em Harish-Chandra Research Institute,
Chhatnag Road, Jhunsi, Allahabad  211019, India}  
}
\normalsize 
\end{center} 
 
\begin{abstract} 
\noindent

Besides supersymmetry, the other prime candidate of physics beyond the
standard model (SM), crying out for verification at the CERN Large Hadron
Collider (LHC), is extra-dimension. To hunt for effects of Kaluza-Klein (KK)
excitations of known fermions and bosons is very much in the agenda of the
LHC. These KK states arise when the SM particles penetrate in the extra
space-like dimension(s).  In this paper, we consider a 5d scenario, called
`Universal Extra Dimension', where the extra space coordinate, compactified on
an orbifold $S^1/Z_2$, is accessed by {\em all} the particles.  The KK number
($n$) is conserved at all tree level vertices. This entails the production of
KK states in pairs and renders the lightest KK particle stable, which leaves
the detector carrying away missing energy.  The splitting between different KK
flavors is controlled by the zero mode masses and the bulk- and brane-induced
one-loop radiative corrections. We concentrate on the production of an $n=1$
KK electroweak gauge boson in association with an $n=1$ KK quark. This leads
to a signal consisting of {\em only one} jet, one or more leptons and missing
$p_T$.  For definiteness we usually choose the inverse radius of
compactification to be $R^{-1} = 500$ GeV, which sets the scale of the lowest
lying KK states. We show on a case-by-case basis (depending on the number of
leptons in the final state) that with 10 ${\rm fb}^{-1}$ integrated luminosity
at the LHC with $\sqrt{s}$ = 14 TeV this signal can be detected over the SM
background by imposing appropriate kinematic cuts.  We record some of the
expectations for a possible intermediate LHC run at $\sqrt{s}$ = 10 TeV and
also exhibit the integrated luminosity required to obtain a 5$\sigma$ signal
as a function of $R^{-1}$.

\vskip 5pt \noindent 
\texttt{PACS Nos:~ 12.60.-i, 04.50.Cd, 13.85.-t } \\ 
\texttt{Key Words:~~Universal Extra Dimension, Kaluza-Klein particles, LHC} 
\end{abstract}

\renewcommand{\thesection}{\Roman{section}} 
\setcounter{footnote}{0} 
\renewcommand{\thefootnote}{\arabic{footnote}} 
 
\section{Introduction} 

Probing the origin of electroweak symmetry breaking (EWSB) constitutes the
prime mandate of the CERN LHC, a proton-proton collider set to operate at
$\sqrt{s}= 14$ TeV.  Unprecedented efforts are going to be invested not only
for the search for the SM Higgs boson, but also for exploring other avenues
which can successfully trigger EWSB.  Among them, supersymmetry and
extra-dimension stand out as two potential rulers of the tera-electron-volt
regime, which is the region of energies and distances to be unplugged at the
LHC.  Although string models are intrinsically extra-dimensional, the
phenomenological implications of extra dimensions were first studied in the
context of a scenario \cite{add} in which gravity propagates in a compact and
flat millimeter size ($1 ~{\rm mm}^{-1} = {10}^{-3}$ eV) extra-dimensional
bulk with the SM particles confined to a 4d brane. The fundamental Planck
scale is then brought down to around a TeV making it accessible to the
collider experiments.  Subsequently, the concept of a `fat-brane' was
introduced \cite{antoniadis} in which the SM particles are not strictly
confined to a point in the extra dimension but travel within the size of the
brane, which may be considered as a bulk of much smaller compactified
dimension. In the present analysis, we attempt to extract possible signals of
such fat-brane scenarios at the LHC, when the size of the brane is order TeV.

Studies based on a scenario in which only the SM gauge bosons access the bulk
while the fermions are confined to a 4d brane \cite{nued} reveal that $R^{-1}$
cannot be below 1-2 TeV. The bound originates from several considerations:
Drell-Yan processes in hadron colliders \cite{Nath:1999mw}, $e^+e^- \to \mu^+
\mu^-$ at LEP2 \cite{Masip:1999mk}, electroweak precision tests
\cite{Rizzo:1999br}, etc. Because the fermions are treated differently from
the bosons, such scenarios are called nonuniversal extra-dimensional models
(NUED).  On the other hand, in what is called the universal extra-dimensional
model (UED) \cite{acd} , where {\em all} the SM particles access the extra
dimension, the constraint is not that tight. UED is relatively easy to
motivate compared to NUED as one does not have to selectively confine the SM
fields in a 4d brane. One crucial difference between UED and NUED is that the
quantized momentum along the extra space direction, conventionally labeled by
the KK number, is conserved for the former but not for the latter.  Thus while
in NUED the KK states mediate processes such as $e^+e^- \to f \bar{f}$ at tree
level incurring very strong experimental constraints, in UED the KK states
appear only in loops resulting in milder bounds. Moreover, one-loop processes
in NUED are ultraviolet divergent while in UED they are finite \cite{db}.
Analysis of constraints on UED from $g-2$ of the muon \cite{nath}, flavor
changing neutral currents \cite{chk,buras,desh}, $Z \to b\bar{b}$ decay
\cite{santa}, the $\rho$ parameter \cite{acd,appel-yee}, hadron collider
studies \cite{collued}, all reveal that $R^{-1}~\gtap~300$ GeV. Consideration
of $b \to s \gamma$, however, implies a somewhat tighter bound
($R^{-1}~\gtap~600$ GeV \cite{bsg}). In fact, UED should be perceived more as
a bare structure with a basic minimum on top of which further details can be
attributed to build separate models for addressing different issues. Several
implications of UED have already been investigated from the perspective of
high energy experiments, phenomenology, string theory, cosmology, and
astrophysics.  To name a few, such TeV scale flat extra-dimensional scenarios
can provide a cosmologically viable dark matter candidate
\cite{Servant:2002aq}, address the issue of fermion mass hierarchy from a
different angle \cite{Arkani-Hamed:1999dc} \footnote{If generation
  universality is assumed in the localization of different fermions, a global
  fit to electroweak observables yields $R^{-1} \gtap~ (2-5)$ TeV. When this
  universality is sacrificed, e.g. in an attempt to address the fermion
  hierarchy problem \cite{Arkani-Hamed:1999dc}, KK gauge bosons pick up tree
  level couplings with different generation of fermions leading to large
  flavor-changing neutral currents and CP violation. Such scenarios cease to
  be of any relevance at the TeV scale, as considerations of $\Delta m_K$,
  $\epsilon_K$ and $\epsilon'_K$ push the lower limit on $R^{-1}$ to around
  5000 TeV \cite{Delgado:1999sv}.}, interpret the Higgs as a quark composite
leading to a successful EWSB without the necessity of a fundamental Yukawa
interaction \cite{Arkani-Hamed:2000hv}, and lower the unification scale down
to a few tens of a TeV \cite{dienes,dienes2,blitz}. In the supersymmetric
context, a new mechanism of supersymmetry breaking related to compactification
has been advanced \cite{antoniadis}, which at the same time ameliorates the
hierarchy problem that an otherwise non-supersymmetric set-up suffers
from. Furthermore, the upper limit of the lightest supersymmetric neutral
Higgs has been shown to be relaxed \cite{relax}. Our present analysis is,
however, based on a non-supersymmetric set-up.

Let us now get into the specifics of UED. There is a single flat extra
dimension ($y$), compactified on an $S^1 /Z_2$ orbifold, and accessed by all
the SM particles \cite{acd}. From a 4d point of view, each field will have an
infinite tower of KK modes, the zero modes being identified as the SM
states. The orbifolding is essential to ensure that fermion zero modes have a
chiral representation. But it has other consequences too. {\em First}, the
physical region along the extra direction $y$ is now smaller [$0, \pi R$] than
the periodicity [$0, 2\pi R$], so the KK number ($n$) is no longer
conserved. What remains actually conserved is the even-ness and odd-ness of
the KK states, ensured through the conservation of KK parity, defined by
$(-1)^n$. {\em Secondly}, Lorentz invariance is also lost due to
compactification, and as a result the KK masses receive bulk and
orbifold-induced radiative corrections \cite{cms1,pk,ggh,viq}. The bulk
corrections are finite and nonzero only for bosons. The orbifold corrections,
which vary logarithmically with the cutoff, depend on group theoretic
invariants, as well as Yukawa and quartic scalar couplings of the gauge and
matter KK fields and hence are flavor-dependent. This induces a mass splitting
among the different flavors of the same KK level, further to what has already
been caused by the different zero mode masses. Typically, if $R^{-1} = 500$
GeV, the lightest among the $n=1$ KK states turns out to be $\gamma^1$
weighing slightly above 500 GeV, just above lie the KK leptons and weak bosons
in the region of 500-550 GeV, further up are the KK quarks near 600 GeV, and
at the peak the KK gluon (the heaviest) hovers around 650 GeV.

Possibility of detection of UED KK states has already been studied in the
context of hadron colliders \cite{nandi}. Two distinct scenarios have been
investigated: (i) the KK states are stable within the size of the detector
(the radiative origin of splitting was not considered here); (ii) although
$\gamma^1$ is the lightest KK state (considering the radiative splitting), the
KK number violating interaction at the brane-bulk interface makes it decay
within the detector to a photon and a graviton (missing particle). In either
of the two options, a lower limit of 350-400 GeV was set on the mass of KK
quarks and gluon from Tevatron Run-I data, while the Run-II data improved the
limit to 500 GeV. It was anticipated that LHC would either discover such
states or at least push the limit to about 3 TeV.  Another approach was
followed in \cite{rizzo}, where the KK states could decay into zero mode
states by KK number violation, and a reach of 3 TeV for KK quarks and gluon
was envisaged with $100^{-1}$ fb luminosity at the LHC.

Although the broad framework within which we work in this paper is the same
UED as pursued in \cite{nandi} and \cite{rizzo}, we differ in some details of
the model.  As a result, our final states are different from theirs, and hence
a comparison of these analyses is not straightforward. Throughout, we strictly
adhere to the conservation of the KK number at all {\em tree level} vertices
and KK parity to all order. This means that our $\gamma^1$ is stable. Then,
once the KK states are produced, the decay patterns and branching ratios are
decided depending on the relative amount of radiative splittings among the
different $n=1$ KK modes. The end products are (at least) two $\gamma^1$,
carrying away missing energy, plus the SM zero modes.  Note,
$\sigma(Q^1\bar{Q^1}) > \sigma(Q^1 {V^1}) > \sigma(V^1 {V^1})$, where $Q^1$
and $V^1$ stand for generic KK quark and KK electroweak gauge boson,
respectively. When both particles at the production vertex are KK
quarks/gluons, although the cross section is very high, nonetheless, the final
states following their decays contain more than one jet making way to very
large SM backgrounds. On the other hand, if both parent KK particles are color
neutral, the final states are hadronically quiet with significantly reduced SM
background but the overall signal production cross section turns out to be
quite low. Therefore, we focus on the production of the mixed combination,
namely $Q^1 V^1$, which is optimally balanced from the signal and background
perspectives. Following the decay chain, we are led to the following final
state configuration: {\em only one} jet, $n_l$ number of leptons ($n_l$ could
be zero to four), and two $\gamma^1$ (missing energy).  This is a path so far
not travelled in the hunt for KK states in a hadronic machine.  As we will
see, for $R^{-1}\sim$ 500 GeV this signal can be comprehensively deciphered
from the SM background with modest integrated luminosities by designing
suitable kinematic cuts.

It is important to mention at this stage that although UED and supersymmetry
are structurally very different theories, ironically, their collider
signatures tantalizingly mimic each other \cite{cms2}. Possible methods of
distinction of UED signals from supersymmetry, mainly based on spin studies,
have been carried out both in the context of the LHC \cite{Smillie:2005ar,
  Datta:2005zs} and the (future) linear collider \cite{Bhattacharyya:2005vm}.
These discriminations require accurate theoretical tools with advanced Monte
Carlos offering high sensitivity to small deviation. In this work, we do not
intend to entertain such multiple possibilities of what might lie at this hazy
domain across the new frontier. Rather, we consider UED as the only new
physics and intend to isolate its signals from possible SM background. Mainly
because of this working hypothesis, a parton-level Monte Carlo that we employ
is good enough for our simulation. We observe that following our strategy KK
states can be spotted at the LHC for $R^{-1}$ all the way up to 700$-$800 GeV
for an integrated luminosity of 100$-$300 $fb^{-1}$.

The paper is organised as follows. In the next section, we briefly introduce
UED, write down the mode expansions of different fields, and give an
estimation of the radiative corrections to different KK masses. In the
subsequent section, we discuss the production of a first level KK quark in
association with the KK gauge bosons and their different decay chains.  Then,
we discuss the relative efficiency of background elimination by looking for
signals with different number of leptons in the final state.  In the last
section, we summarize and conclude.

\section{Universal Extra Dimension}
\subsection{Mode Expansions} 

The extra coordinate $y$ is compactified on a circle of radius $R$ with a
$Z_2$ orbifolding identifying $y$ with $-y$. The orbifolding is essential for
generating zero mode {\em chiral} fermions. After the $y$-dependence is
integrated out, the 4d Lagrangian contains the zero mode and the KK modes of
different fields.  Let us now take a look at the KK mode expansions of these
fields. Since $Z_2$ is a symmetry of the theory, each 5d field must be either
even or odd under this discrete parity. The KK expansions are given by,
\begin{eqnarray}
\label{fourier}
A_{\mu}(x,y)&=&\frac{\sqrt{2}}{\sqrt{2\pi
R}}A_{\mu}^{0}(x)+\frac{2}{\sqrt{2\pi
R}}\sum^{\infty}_{n=1}A_{\mu}^{n}(x)\cos\frac{ny}{R},~~~~
A_5(x,y) = \frac{2}{\sqrt{2\pi
R}}\sum^{\infty}_{n=1}A_5^{n}(x)\sin\frac{ny}{R}, \nonumber\\
\phi(x,y)&=&\frac{\sqrt{2}}{\sqrt{2\pi
R}}\phi^{0}(x)+\frac{2}{\sqrt{2\pi
R}}\sum^{\infty}_{n=1}\phi^{n}(x)\cos\frac{ny}{R}, \nonumber \\
\mathcal{Q}_{i}(x,y)&=&\frac{\sqrt{2}}{\sqrt{2\pi 
R}}\bigg[{\pmatrix{u_i\cr d_i}}_{L}(x)+\sqrt{2}\sum^{\infty}_{n=1}\Big[
\mathcal{Q}^{n}_{iL}(x)\cos\frac{ny}{R}+
\mathcal{Q}^{n}_{iR}(x)\sin\frac{ny}{R}\Big]\bigg], \\
\mathcal{U}_{i}(x,y)&=&\frac{\sqrt{2}}{\sqrt{2\pi
R}}\bigg[u_{iR}(x)+\sqrt{2}\sum^{\infty}_{n=1}\Big[
\mathcal{U}^{n}_{iR}(x)\cos\frac{ny}{R}+
\mathcal{U}^{n}_{iL}(x)\sin\frac{ny}{R}\Big]\bigg], \nonumber\\
\mathcal{D}_{i}(x,y)&=&\frac{\sqrt{2}}{\sqrt{2\pi
R}}\bigg[d_{iR}(x)+\sqrt{2}\sum^{\infty}_{n=1}\Big[
\mathcal{D}^{n}_{iR}(x)\cos\frac{ny}{R}+
\mathcal{D}^{n}_{iL}(x)\sin\frac{ny}{R}\Big]\bigg], \nonumber
\end{eqnarray}
where $i=1,2,3$ are generation indices.  The complex scalar field $\phi (x,y)$
and the gauge boson $A_\mu(x,y)$ are $Z_2$-even fields, and their zero modes
are identified with the SM scalar and gauge boson respectively. The field
$A_5(x,y)$ is a real scalar transforming in the adjoint representation of the
gauge group, and it does not have any zero mode. The fields $\mathcal{Q}$,
$\mathcal{U}$, and $\mathcal{D}$ describe the 5d quark doublet and singlet
states, respectively, whose zero modes correspond to the chiral SM quark
states. The mode expansions of the doublet and singlet leptons can be written
{\em mutatis mutandis}.

\subsection{Radiative corrections to the KK masses}
At a given $n$, the KK mass is given by $\sqrt{m_0^2 + n^2/R^2}$. So, {\em
  modulo} the zero mode masses, the KK states are degenerate. But this is only
a tree level result. Radiative corrections lift this degeneracy
\cite{cms1,pk,ggh,viq}. To provide intuition, let us consider the kinetic term
of a scalar field as \cite{cms1} $L_{\rm kin} = Z \partial_\mu
\phi \partial^\mu \phi - Z_5 \partial_5 \phi
\partial^5 \phi ~~(\mu = 0,1,2,3)$, where $Z$ and $Z_5$ are renormalization
constants. Recall that tree level KK masses originate from the kinetic term in
the $y$-direction. If $Z = Z_5$, there is no correction to those masses. But
this equality follows from Lorentz invariance. When a direction is
compactified, Lorentz invariance breaks down.  Then $Z \neq Z_5$, leading to
$\Delta m_n \propto (Z-Z_5)$. More specifically, there are two kinds of
radiative corrections.

\begin{table}[tbh]
\begin{center}
\begin{tabular}{|c|c|c|c|c|c|c|c|c|c|}  \hline 
States & ${Q^1}$ & ${u^1}$ & ${d^1}$ & ${L^1}$ & ${e^1}$  & 
${g^1}$ & ${{W^{\pm}}^1}$ & ${Z^1}$ & ${\gamma ^1}$ \\ 
\hline
Mass & 598.7 & 587.3 & 585.5 & 515 & 505.5 & 642.3 & 536 & 542.1 & 501.0 \\ 
\hline 
$a$ & 3 & 3 & 3 & 0 & 0 & 23/2  & 0 & 0 & 0 \\
\hline
$b$ & 27/16 & 0 & 0 & 27/16 & 0  & 0  & 15/2  & 15/2  & 0 \\
\hline
$c$ & 1/16  & 1  & 1/4  & 9/16  & 9/4  & 0  & 0  & 0  & $-$1/6\\
\hline
\end{tabular}
\caption{\sf \small{Radiatively corrected first KK-mode masses (all in GeV)
    for $R^{-1} =$ 500 GeV and $\Lambda R = 20$. The values of $a,b,c$,
    introduced in Eq.~(\ref{abc}), are displayed. While assigning $b$ and $c$
    for $\gamma^1$ and $Z^1$, we used $\theta_W^1 \to 0$.}} 
\label{kkmass}
\end{center}
\end{table}

{\em Bulk correction}:~ These corrections are finite and nonzero only for
bosons. They arise whenever the internal loops wind around the compactified
direction. These corrections, for a given field, are the same for any KK mode.
For a KK boson mass $m_n(B)$, these corrections are given by
\begin{eqnarray}
\label{kappa}
\delta\, m_n^2(B) &=&  \kappa \frac{\zeta(3)}{16\pi^4} 
\left(\frac{1}{R}\right)^2,
\end{eqnarray}
where $\kappa$ is a collective representation of group invariants, being equal
to $-39 g_1^2/2, -5 g_2^2/2$ and $-3 g_3^2/2$ for $B^n, W^n$ and $g^n$,
respectively. Clearly, for $R \rightarrow \infty$, one recovers the original
Lorentz invariance and the correction vanishes.

{\em Orbifold correction}:~ Orbifolding additionally breaks translational
invariance in the $y$-direction. The corrections to the KK masses arising from
interactions localized at the fixed points are logarithmically divergent. The
corrections can be thought of as counterterms whose finite parts are
unknown. We just follow a predictive hypothesis that these corrections vanish
at the cutoff scale $\Lambda$. The amount of this correction to a generic KK
fermion mass $m_n(f)$, or a KK gauge boson mass $m_n(B)$, is given by
\begin{eqnarray}
\label{abc}
\frac{\delta m_n(f)}{m_n(f)} \left(\frac{\delta m_n^2(B)}{m_n^2(B)} \right)
& = & \left(a \, \frac{g_3^2}{16\pi^2}
+ b \,\frac{g_2^2}{16\pi^2} + c \,\frac{g_1^2}{16\pi^2}
\right) \,\ln\frac{\Lambda^2}{\mu^2},
\end{eqnarray}
where $a,b$ and $c$ for different KK states are listed in Table
\ref{kkmass}.  As it turns out, the orbifold corrections are
numerically more significant than the bulk corrections.

The mass squared matrix of the neutral KK gauge boson sector in the 
$B_n$, $W^3_n$ basis is given by
\begin{equation}
\label{masssq}
\left( 
\begin{array}{cc}
\frac{n^2}{R^2}+ \hat{\delta} m_{B_n}^2 + \frac{1}{4}g_1^2 v^2 
& \frac{1}{4}g_1 g_2 v^2 \\
\frac{1}{4}g_1 g_2 v^2 & \frac{n^2}{R^2}+ \hat{\delta} m_{W_n}^2 +
\frac{1}{4}g_2^2 v^2
\end{array}
\right),
\end{equation}
where $\hat{\delta}$ represents the sum of bulk and orbifold radiative
corrections. The KK photon and $Z$ boson states are obtained by diagonalizing
the above matrix. Note, the value of the weak mixing angle for the KK states
is sizably altered from the zero mode value ($\sin^2 \theta_W \simeq 0.23$)
due to a difference in size of those $\hat\delta$-terms in the mass squared
matrix in Eq.~(\ref{masssq}). The modified value is different for different
choices of $n$ and $R$. For $n=1$ and $R^{-1} = 500$ GeV, it turns out that
$\sin^2 \theta_W^1 \sim 0.01$, i.e., $\gamma^1$ and $Z^1$ are primarily $B^1$
and $W_3^1$, respectively.

In Table \ref{kkmass}, we present the masses for different $n=1$ KK
excitations to assess the extent to which the radiative corrections lift the
degeneracy. For illustration, we take $R^{-1} = 500$ GeV and $\Lambda R =
20$ (a rough justification for this choice is that the gauge couplings,
following a power law renormalization group running, tend to unify after more
or less $\Lambda R = 20 - 25$ KK resonances are excited).

\section{Productions and decay of the first KK-mode}\label{production}
\begin{table}
\begin{center}
\begin{tabular}{|c|c|c|c|c|}
\hline 
Excited quark $\rightarrow$&\multicolumn{2}{c|}{SU(2) Doublet
($Q$)}&\multicolumn{2}{c|} {SU(2) Singlet ($q$)} \\
\hline
Excited boson $\downarrow$&$\;\;\;\;\;\;\;\;\;a_R\;\;\;\;\;\;\;\;\;\;\;$ & 
$\;\;\;\;\;\;\;\;\;a_L\;\;\;\;\;\;\;\;\;\;\;$ &
$\;\;\;\;\;\;\;\;\;a_R\;\;\;\;\;\;\;\;\;\;\;$ &
$\;\;\;\;\;\;\;\;\;a_L\;\;\;\;\;\;\;\;\;\;\;$\\
\hline 
$W^{1}$ &$0$ &$\frac{g}{\sqrt{2}}$&$0$&$0$ \\
$Z^{1}$ &$0$ &$\frac{g}{2\cos\theta_W^1}\left(T_3 - e_Q \sin^2
\theta_W^1 \right)$
&$-\frac{g}{2\cos\theta_W^1}\left(e_q \sin ^2
\theta_W^1\right)$&$0$ \\
$\gamma^{1}$ &$0$ &$\frac{e_Q }
{\cos\theta_W}\cos \theta_W^1$&$\frac{e_q }
{\cos\theta_W}\cos \theta_W^1$&$0$ \\
\hline
\end{tabular}
\caption{\sf \small {The couplings $a_L$ and $a_R$ -- involving an excited
  gauge boson, an excited quark, and an ordinary quark -- used in
  Eq.~(\ref{matel}). Note that KK-parity conservation requires $a_R$ ($a_L$)
  to vanish for $Q$ ($q$) in all cases. The couplings of the excited leptons
  ($L^1, l^1$) to the excited gauge bosons follow a similar pattern and can be
  easily read off from this Table.}}
\label{coupl}
\end{center}
\end{table}

As noted in the previous section, if $R^{-1}$ is not too large, the first
KK-excitations of the Standard Model particles are in the right mass range for
pair-production at the LHC. We consider the parton-level process\footnote{Here
  $\overline{Q}$ stands for the SU(2) doublet ($Q^1$) as well as the singlet
  ($q^1 \equiv u^1, d^1$) KK excited quarks.}  $qg \rightarrow \overline{Q}
V^1$ for which the matrix element square is:
\begin{eqnarray}
|\mathcal{M}\{qg \rightarrow \overline{Q} V^1\}|^2 =
\frac{\pi \alpha_s(\hat s) (a_L^2 + a_R^2)}{6} 
\left[{{\{-2\hat s\hat t + 2\hat sm^2_{\overline{Q}}\}}\over {{\hat s}^2}}
+{\{- 2\hat s\hat t -  4\hat tm^2_{\overline{Q}} + 2\hat sm^2_{\overline{Q}} +
4m_{V^1}^2m^2_{\overline{Q}}\}\over{(\hat t - m^2_{\overline{Q}})^2}} \right. 
\nonumber \\
\left. + {2\{-2\hat tm^2_{\overline{Q}} + 2(\hat s + \hat t)m_{V^1}^2 +
2m_{V^1}^2m^2_{\overline{Q}} - 2m_{V^1}^4\}\over {\hat s(\hat t -
m^2_{\overline{Q}})} }\right],
\label{matel}
\end{eqnarray}
where $\hat s$ and $\hat t$ are the (parton-level) Mandelstam variables. The
couplings $a_L$ and $a_R$ are fixed by the final state particles and are
summarized in Table \ref{coupl}.  They depend on the weak mixing angle of the
excited bosons, $\theta_W^1$, which is a function of $R^{-1}$ and is
considerably smaller than $\theta_W$. The KK-excitations of quarks and leptons
are vector-like fermions, so, unlike for the SM fermions, the couplings listed
for `singlet' or `doublet' quarks have both chiral counterparts.
\begin{center}
\begin{figure}[thb]
\hskip 1.0cm
\includegraphics[width=15cm,height=6.5cm,angle=0]{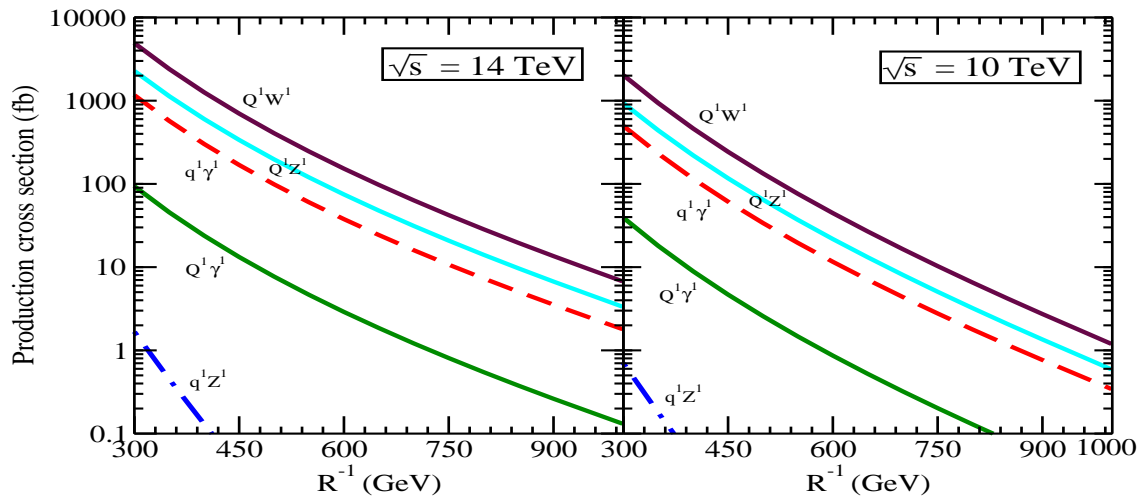}
\caption{\sf \small {Cross sections for the associated
production of the lightest KK electroweak gauge bosons with the
lightest KK
quarks. Solid (broken) lines correspond to  SU(2) doublet quarks
$Q^1$ (SU(2) singlet quarks $q^1$).
Note that the $W^1$ cannot be
produced with a $q^1$.  }}
\label{f:cross}
\end{figure}
\end{center}
\vskip -1.2cm

The cross sections for the various channels at the LHC with $\sqrt{s}$ = 14
TeV and 10 TeV (evaluated using CTEQ6 parton distributions (leading order
parametrization) \cite{cteq} with $Q^2 = \hat s$)\footnote{We have performed
  two independent checks to assess the QCD uncertainties.  The MRST parton
  distribution function, we have checked, yields cross sections which at the
  leading order are within 10\% of the CTEQ6L1 results. We also varied the
  scale $Q^2$, which we kept common for both parton density and $\alpha_S$,
  from $2\hat{s}$ to $0.5\hat{s}$, and observed that the leading order cross
  sections alter by about (7-8)\% around the values quoted for $Q^2 =
  \hat{s}$.}  are shown as a function of $R^{-1}$ in Fig.~\ref{f:cross}.  The
cross sections fall rapidly with $R^{-1}$ and vary over several orders of
magnitude, $q^1Z^1$ having the smallest cross section and $Q^1W^1$ the largest
(in $pb$ order for $R^{-1} \sim 500$ GeV)\footnote{Our cross sections are in
  almost complete agreement with the ones obtained using MADGRAPH
  \cite{madgraph} and FEYNRULES \cite{feynrules} (more precisely, FR-MUED).}.

We next turn to the expected final state event topologies. These
can be ascertained from the decay characteristics of the excited
quarks and gauge bosons. The possible decay channels and
branching ratios have been listed in \cite{cms2}. For the discussions
below, it is worth noting that $\sin^2\theta^1_W \ll
\sin^2\theta_W$.
\begin{itemize}

\item Excited $\gamma$ ($\gamma^1$): $\gamma^1$ is the lightest Kaluza-Klein
  particle (LKP).  This particle is uncharged and stable due to KK-parity
  conservation. When produced, it escapes detection.

\item Excited $Z$ ($Z^1$): The $Z^1$ being lighter than the excited quark
  states cannot decay to them. Nor is the decay to a $Z$ and $\gamma^1$
  kinematically allowed. So, a $Z^1$ will decay to the final
  states\footnote{Recall that $L$ ($l$) represents the SU(2) doublet (singlet)
    charged lepton state.}  $(L^1)^\pm L^\mp$ and $\nu^1 \nu$ with equal
  branching ratios\footnote{The decay $Z^1 \rightarrow (l^1)^\pm l^\mp$ is
    suppressed by $\sin^2\theta^1_W$.}.  In the former case there is the
  subsequent decay $L^1 \rightarrow L \gamma^1$ of which only the lepton is
  observable and the $\gamma^1$ remains undetected.  This leads to a final
  $Z^1$ signal consisting of two (oppositely) charged leptons of the same
  flavor and missing $p_T$.

\item Excited $W$ ($W^1$): Kinematics does not permit the $(W^1)^\pm$ to decay
  hadronically or to $W^\pm \gamma^1$.  Consequently, it decays to either
  $L^\pm \nu^1$ or to $(L^1)^\pm \nu$ with equal branching ratios. The final
  decay products are therefore $L^\pm \nu \gamma^1$ of which only the charged
  lepton is observable.

\item SU(2) singlet excited quarks ($q^1$): The SU(2) singlet excited quark
  decays predominantly to a zero-mode quark and a $\gamma^1$. The coupling to
  $Z^1$ is suppressed by $\sin^2\theta^1_W$.

\item SU(2) doublet excited quarks ($Q^1$): These can decay to a zero-mode
  quark doublet $Q$ and any of the excited gauge bosons $W^1, Z^1,
  \gamma^1$. However, the dominant decay modes turn out to be \cite{cms2} the
  first two modes with a 2:1 ratio.
\end{itemize}

The zero-mode quarks and leptons produced from the decay of different excited
gauge bosons and quarks are listed in Table \ref{dkql}. In addition to the
particles shown, there is always a $\gamma^1$ in the final state carrying away
missing $p_T$. As a consequence, $(Q^1 V^1)/(q^1 V^1)$ production results in
a final state containing 1-jet + $n_l$ leptons + missing $p_T$. Depending on
the decay modes of the excited quark and $V^1$, the number of leptons, $n_l$,
can be 0 to 4. In the following, we will classify the signal according to the
number of detected leptons and compare it with the SM background.
\begin{table}[thb]
\begin{center}
\begin{tabular}{|c|c|c|c|c|}
\hline 
Parent particle &$Z^1$&$(W^1)^\pm$
&$q^1$&$Q^1$ \\
\hline
Decay products&$(L^\pm + L^\mp)$ & 
$L^\pm + \nu$ & $q$ & $Q$ + $(L^\pm + L^\mp)$ or $(\nu \nu)$\\
  &or $(\nu \nu)$ &  & & or $Q'$ +  $(L^\pm + \nu)$\\

\hline
\end{tabular}
\caption{\sf The zero-mode quarks or leptons produced from the
decays of $Z^1, (W^1)^\pm, q^1,$ and $Q^1$. In addition, a
$\gamma^1$ is also produced in all decays. }
\label{dkql}
\end{center}
\end{table}

In our analysis we include contributions from both electron and muon final
states. Tau-leptons will also be produced at essentially the same
rates. Conservatively, we have not included the states arising from their
purely leptonic decays (branching ratio $\sim$ 0.17 to electrons or muons)
which would produce $e^+e^-$, $\mu^+\mu^-$, and $e^\pm \mu^\mp$ events in the
ratio 1:1:2.

\section{Multilepton signals and backgrounds}\label{simulation}
In this section, one by one we consider the different multilepton final states
which may arise from $(Q^1 V^1)/(q^1 V^1)$ production.  Table \ref{dkql} will
be of use for the discussions below.  As stressed earlier, in addition to the
leptons, there is always one hadronic jet and missing transverse momentum,
$\not \!\!{p_T}$, in the signal.

{\em Basic cuts}:~ For the signal and the background a minimal cut of
${p_T}^{\rm jet} > 20$ GeV has been applied while leptons are required to
satisfy $p_T >$ 5 GeV.  A cut of $\not \!\!{p_T} > $ 25 GeV has also been
used. To evade a possible large background from lepton pairs produced by soft
photons we require $M_{l_il_j} >$ 5 GeV in multilepton events.  A rapidity cut
of $|\eta| <$ 2.5 is applied for leptons and the jet.  We call these the
`basic' cuts.

Numerical values of the cross sections after the basic cuts are listed in the
Table \ref{cross_tab} for two different values of $R^{-1}$ along with the
backgrounds. We have estimated the SM backgrounds at the partonic level using
MADGRAPH.  CTEQ6 parton distribution functions are utilised.  It is seen that
the backgrounds far overwhelm the signals in all cases but for the 4-lepton
one. Further kinematic cuts, discussed below, are needed to enhance the signal
{\em vis-\`{a}-vis} the background.

\begin{table}[tbh]
\begin{center}
  \begin{tabular}{|c|c|c|c|c|c|} \hline Channel & 0$l$ & 1$l$ & 2$l$ & 3$l$ &
    4$l$ \\ \hline Signal (500 GeV) & 106.4 & 17.92 & 29.58 & 9.39 & 1.01 \\
    \hline Signal (1 TeV) & 2.02& 0.35 & 0.606 & 0.210 & 0.025 \\ \hline
    Background & 4.7$\times 10^5$ & 1.3$\times 10^6$ & 8.6$\times 10^4$ &
    1183.21 & 0.13 \\ \hline
\end{tabular}

\caption{\sf \small{ Cross section (in $fb$) for multilepton channels of the
    signal (for $R^{-1} =$ 500 GeV and 1 TeV) and SM background at the LHC
    with $\sqrt{s}$ = 14 TeV after the basic cuts. The two-lepton number
    corresponds to $e^+e^-$ or $\mu^+\mu^-$.}}
\label{cross_tab}
\end{center}
\end{table}

\subsection{No lepton}
This case will result from the production of (i) $q^1 \gamma^1$ and (ii) $q^1
Z^1$ followed by an invisible decay ($\nu \nu \gamma^1$) of the $Z^1$.  It is
seen from Table \ref{cross_tab} that the signal cross section after the basic
cuts is rather large compared to the other cases since there is no branching
ratio (to lepton) suppression incurred here. This is unfortunately offset by
the very huge SM background in this channel.  In this work our interest is to
use the multilepton final state to reduce the SM background. So, this
no-lepton topology is not pursued any further.

\subsection{One lepton}
This final state can arise from (i) $Q^{1} Z^{1}$ (and $Q^{1} \gamma^{1}$)
production followed by the decay $Q^{1} \rightarrow Q^{\prime} W^{1}$ and an
invisible $Z^{1}$ decay, and (ii) $Q^{1} W^{1}$ followed by $Q^{1} \rightarrow
Q Z^{1}$ and an invisible $Z^{1}$ decay. Whenever there are multiple modes
which can contribute to a signal, we have included all of them together in the
analysis.

The SM background to this one-lepton signal comes from $W$-production in
association with a jet, followed by leptonic decay of the $W$-boson. The rate
of the irreducible one-lepton background at LHC energies is large compared to
the signal cross section. Application of kinematical cuts is not sufficient to
dig the signal out from this background.  Unraveling the one-lepton signal
over the SM background could be too challenging.

\subsection{Two leptons}
Next we consider the two-lepton case. The signal can arise from (i)
$Q^{1}W^{1}$ production followed by $Q^{1} \rightarrow Q^{\prime} W^{1}$ and
(ii) $Q^{1} Z^{1}$ production followed by the decays : $Q^{1} \rightarrow Q
Z^{1}$ and the invisible decay of one of the two $Z^1$. Note that mode (i) can
lead to $e^\pm \mu^\mp$ final states.  We separately consider `like-flavor',
i.e., $e^+e^-$ or $\mu^+\mu^-$, as well as `unlike-flavor', i.e., $\mu^+e^-$ +
$e^+\mu^-$, in the following.

The dominant SM background to this final state will be from $t\bar{t}$ and
$b\bar{b}$ production followed by their semileptonic decay. In addition, the
electroweak production of $t\bar{b}$, $b\bar{t}$ and their subsequent decay is
also of significance. These backgrounds are severely cut down by requiring the
monojet to satisfy $p_T^{\rm jet} > 20$ GeV and the other basic cuts.

From top-pair production and semileptonic decays of both, two jets are
expected. To mimic the signal, one of these jets must fail the jet-$p_T$ cut
while the other must pass. In addition, we demand `lepton isolation' -- that
all leptons be isolated from the jet satisfying $\Delta R >$ 0.7, where
$\Delta R^2 = (\Delta \phi^2 + \Delta \eta^2)$. This and the other basic cuts
reduce the cross section to 173.2 (79.84) $fb$ for an LHC run with $\sqrt{s}$
= 14 (10) TeV.  Further kinematic cuts are necessary, as discussed below, to
reduce this background.

The $b\bar{b}$ production followed by semileptonic decays has a large cross
section of $\sim$ 100 $nb$ at the LHC. However, the low mass of the $b$-quark
ensures that this background is {\em totally} removed when the basic cuts are
imposed along with lepton isolation.

In spite of their electroweak origin, the $t\bar{b}$, $b\bar{t}$ production
rates are substantial.  The twin requirements of one jet with ${p_T}^{\rm jet}
> 20$ GeV and a lepton isolation cut of $\Delta R >$ 0.7 are found to suffice
for eliminating this background. After these cuts, this channel contributes
15.94 (10.92) $fb$ to the background for $\sqrt{s}$ = 14 (10) TeV.

The remaining SM background is from $W$ pair production in association with a
jet. $Z$ pair (real or virtual) or $Z\gamma^\ast$ (leptons coming from the
$\gamma^\ast$ with an invisible $Z$ decay) production in association with a
jet also contributes to the SM background.  The $W$ pair production channel,
which is the more relevant one since the on-shell $Z$ background is small and
anyway readily removed, results in $e^+e^-$, $\mu^+\mu^-$, and $e^\pm \mu^\mp$
events in the ratio 1:1:2. In these cases, the jet is from either gluon or
quark radiation off the initial partons, and consequently, most of the time it
emerges close to the beam axis. This will be reflected in the rapidity
distribution of the jet. This is in contrast with the signal. The basic cut on
${p_T}^{\rm jet}$ helps enhance the strength of the signal over the SM
background.

Here, we focus on the case of $\mu^+\mu^-$ + one jet + missing $p_T$ final
states. The results are identical if the $\mu^+\mu^-$ are replaced by
$e^+e^-$. Parallely, we will be remarking on the $\mu^\pm e^\mp$ alternative.
For $R^{-1}$ = 500 GeV, with the LHC running at $\sqrt{s}$ = 14 TeV the signal
turns out to be much less than the background in both cases (see Table
\ref{tab:2L}).  Naturally, for $\sqrt{s}$ = 10 TeV the situation is worse.  To
enhance the signal {\em vis-\`{a}-vis} the background the following further
kinematic cuts are
suggested\footnote{$l_1$ has lower $p_T$ than $l_2$.}:\\

\centerline{
(i)  ${p_T}^{l_1} <$ 25 GeV,  
(ii)  ${p_T}^{l_2} <$ 25 GeV, and 
(iii) $|M_{l_1l_2} - M_Z| >$ 10 GeV}.

\begin{center}
\begin{figure}[tbh]
\vskip -1.0cm
\includegraphics[width=18cm,height=18cm,angle=0]{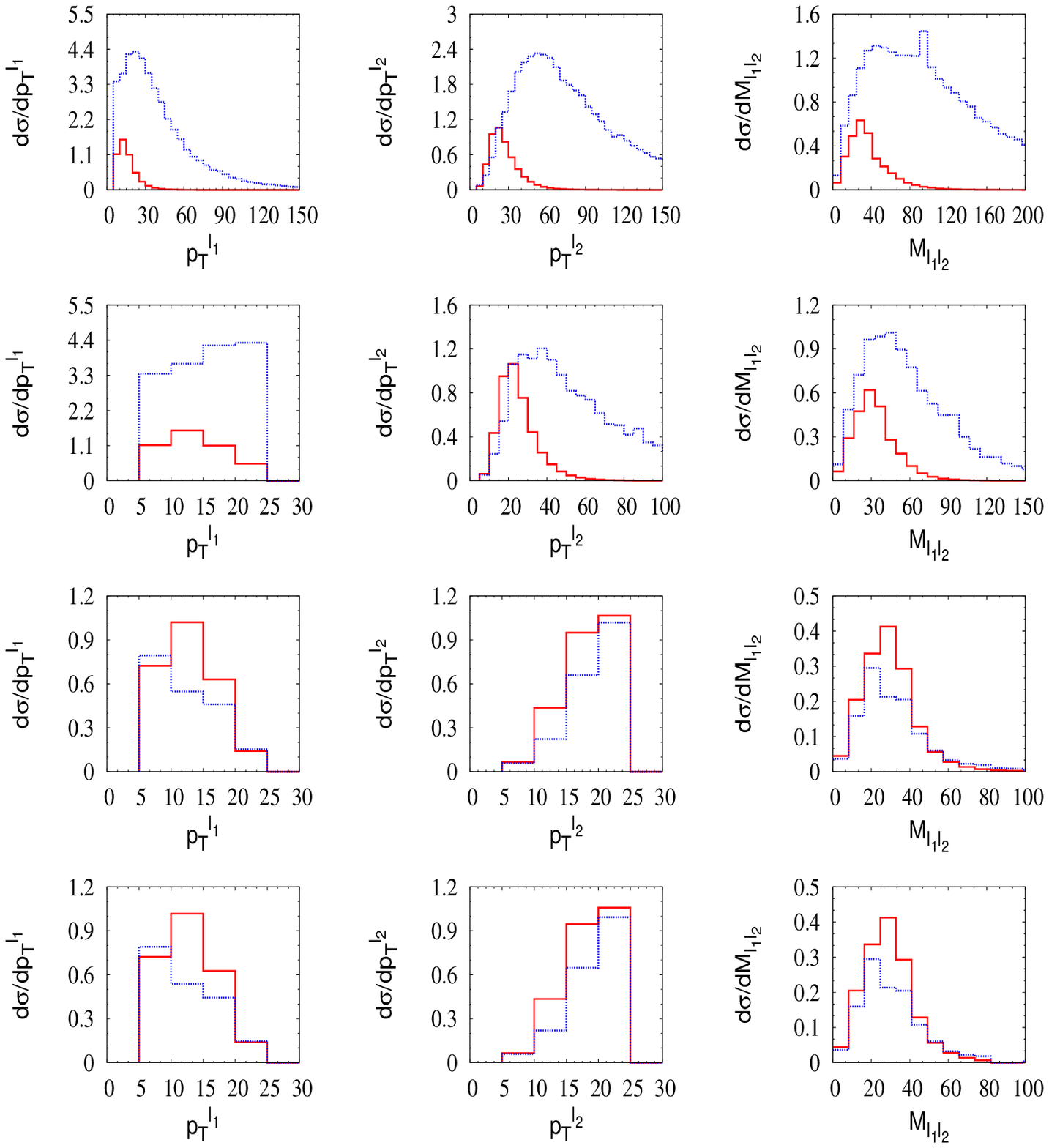}
\vskip -6.0cm
\caption{\sf \small Two like-flavor leptons ($\mu^+\mu^-$ or $e^+
e^-$) + jet + missing $p_T$ signal (for
$R^{-1}$ = 500 GeV) 
and background differential cross sections (in $fb$/GeV). The solid
(red) histograms are for the signal and the dotted (blue) ones
are for the background. The successive rows reflect the impact of
kinematic cuts shown in Table \ref{tab:2L}. Lepton isolation cut has already
been imposed on the top row.} 
\label{fig:2L}
\end{figure}
\end{center}

The effect of these cuts on the signal (red solid histogram) and the
background (blue dotted histogram) are shown in  Fig.~\ref{fig:2L} and
also presented in Table \ref{tab:2L}.

\begin{table}[tbh]
\begin{center}
\begin{tabular}{|c|c|c|c|c|}  \hline 
$\sqrt{s}\;\;\;\; \rightarrow$ &\multicolumn{2}{|c|}{14 TeV}&
\multicolumn{2}{|c|}{10 TeV} \\ \hline
Cut used   $ \downarrow$          & Signal       & Background  &
Signal & Background \\  \hline

Basic cuts            & 29.58 (43.10)    & 8.6$\times 10^4$
(17.2$\times 10^4$)  & 10.00 (14.60) & 4.8$\times 10^4$ (9.6$\times
10^4$)\\ \hline
Lepton isolation            & 24.24 (35.24)    & 218.38 (429.64)  &
8.28 (12.06) & 108.54 (212.78)\\ \hline 
${p_T}^{l_1} < 25$ GeV  & 21.66 (30.88) & 78.67 (154.90)  & 7.52
(10.74) & 41.10 (80.70) \\ \hline
${p_T}^{l_2} < 25$ GeV  & 12.58 (18.00)  & 9.44 (18.40) & 4.53
(6.52) & 5.27 (10.22) \\ \hline
$|M_{l_1l_2} - M_Z| >$ 10  GeV & 12.52 (17.88) & 9.18 (17.98) & 4.51 (6.48) 
 & 5.17 (10.08) \\ \hline  
\end{tabular}
\caption{\sf \small{Cross section (in $fb$) at the LHC   of
signal and background for the like-flavor $\mu^+\mu^-$ or
$e^+e^-$ (unlike-flavor $\mu^+e^-$ + $e^+\mu^-$) dilepton plus
one jet and missing $p_T$ channel for $R^{-1}$ = 500 GeV.}}
\label{tab:2L}
\end{center}
\end{table}

The two cuts on the lepton $p_T$ are chosen on the following grounds.  In UED,
the leptons arising from the decays (at any stage of the decay chain) of
KK-mode excitations are always accompanied by the LKP or some other KK
excitation.  The small mass splitting between the KK excitations results in a
comparatively soft lepton and their $p_T$ distributions will be peaked around
lower values and are spread over a limited range. For the SM background, the
energy of the parent particles ($Z$, $W$ or $\gamma^\ast$), which are much
lighter than the KK states, is shared between two particles of negligible
mass. As a result, $p_T$ distributions of these leptons, though also peaked at
lower values, have a tail extended to higher values compared to the signal.
So, by demanding the $p_T$ of the leptons to be bounded from above, one can
get rid of much of the background. This is exemplified in the top three rows
of Fig.~\ref{fig:2L}.

As is seen from Table \ref{tab:2L}, for the $\sqrt{s}$ = 14 TeV case, through
the kinematic cuts the background for like- (unlike-)flavor is reduced from
8.6$\times 10^4$ (17.2$\times 10^4$)$fb$ to 9.18 (17.98)$fb$. For a modest
integrated luminosity of 10$fb^{-1}$ the significance\footnote{Using the above
  definition of significance is quite appropriate especially since in some
  cases we are dealing with very low backgrounds following the imposition of
  the kinematic cuts.} ($S/\sqrt{S+B}$) after the basic cuts is 0.32 (0.33),
which after the kinematic cuts is enhanced to a healthy 8.50 (9.44).

\begin{figure}[thb]
\vskip 1.5cm
\hskip 2cm
\includegraphics[width=12cm,height=6cm,angle=0]{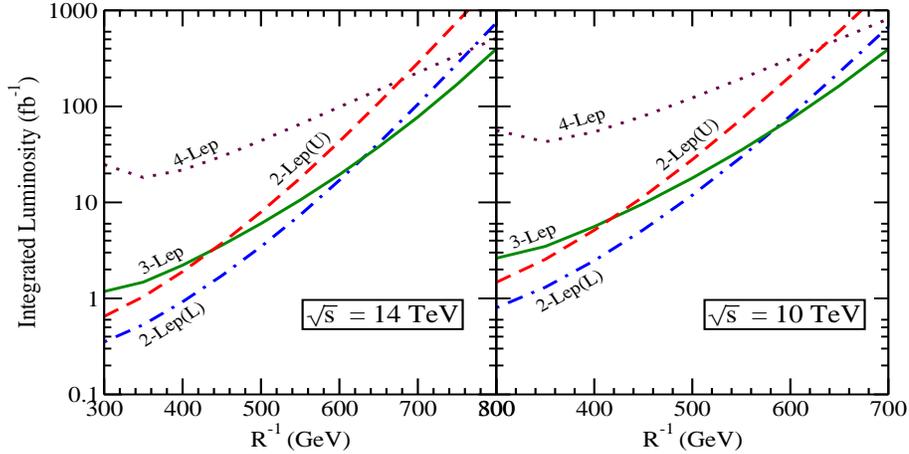}
\caption{\sf \small {The required integrated luminosity at the
LHC running at $\sqrt{s}$ = 14 TeV (left panel) and 10 TeV (right
panel) for a 5$\sigma$ signal in the multilepton + 1 jet +
missing $p_T$ channels as a function of $R^{-1}$.  Results are
shown for 2-, 3-, and 4-leptons.  `U' and `L' correspond to the
cases of unlike- and like-lepton flavors.  }}
\label{lumilim}
\end{figure}

For the discussion so far we have chosen as a reference value $R^{-1}$ = 500
GeV.  The integrated luminosities necessary for a 5$\sigma$ signal after the
complete set of kinematic cuts as a function of $R^{-1}$ for $\sqrt{s}$ = 14
TeV (left) and 10 TeV (right) are plotted in Fig.~\ref{lumilim}. Both
like-flavor (L) $\mu^+\mu^-$ or $e^+ e^-$ (blue dot-dashed) and unlike-flavor
(U) $\mu^+e^-$ + $e^+\mu^-$ (red-dashed) cases are shown.  It is seen that
with an integrated luminosity of 100$fb^{-1}$ for the like-flavor case one can
probe the KK states for $R^{-1}$ upto about 700 (600) GeV for the LHC running
with $\sqrt{s}$ = 14 (10) TeV while for the unlike-flavor case these limits
are 620 (570) GeV.

\subsection{Three leptons}
We now turn to the three-lepton case. As the number of leptons in the final
state increases the signal gains over the background. For the signal, channels
with different numbers of leptons follow from alternate decay modes of the
produced KK particles and the cross sections differ only to the extent of the
corresponding branching ratios. On the other hand, a background channel with
more leptons usually corresponds to a higher order electroweak process with
its concomitant perturbative suppression. Alternatively, it involves a
$Z$-boson decay which can be readily removed by an invariant mass cut. This
will be brought out in the three- and four-lepton channels to which we now
turn.

The three-lepton final state is realised through (i) $Q^{1}W^{1}$ production
followed by $Q^{1} \rightarrow Q^{0} Z^{1}$ and (ii) $Q^{1} Z^{1}$ production
followed by the decays: $Q^{1} \rightarrow Q^{\prime 0} W^{1}$ and $Z^{1}$
decay producing two leptons with $\gamma^1$ (in two steps).  The SM background
for the three-lepton plus jet and missing $p_T$ final state will arise from
$t\bar{t}$ production, $WZ$ or $W\gamma^\ast$ production in association with a
jet.

The first step in enhancing the signal compared to the background is to apply
the jet-lepton isolation cut ($\Delta R >$ 0.7) on every lepton.  For $R^{-1}$
= 500 GeV, the three-lepton background that still survives turns out to be
about three times the signal. We order the three leptons in increasing
${p_T}$: ${p_T}^{l_1} < {p_T}^{l_2} < {p_T}^{l_3}$ and apply additional cuts
to enhance the signal compared to the background.

\begin{table}[tbh]
\begin{center}
\begin{tabular}{|c|c|c|c|c|}  \hline 
$\sqrt{s}\;\;\;\; \rightarrow$ &\multicolumn{2}{|c|}{14 TeV}&
\multicolumn{2}{|c|}{10 TeV} \\ \hline
Cut used   $ \downarrow$          & Signal       & Background  &
Signal & Background \\  \hline
Basic cuts            &  9.39   & 1183.21   & 3.21  & 555.85 \\ \hline 
Lepton isolation            &  6.96   & 21.69   & 2.41  & 10.53 \\ \hline 
${p_T}^{l_2} < 25$ GeV  &  5.63   & 4.09   & 2.01 & 1.75 \\ \hline
${p_T}^{l_3} < 40$ GeV  & 5.12    & 1.31  & 1.86 & 0.64 \\ \hline
$|M_{l_il_j} - M_Z| >$ 10  GeV & 5.03  & 1.16 & 1.82 & 0.57 \\ \hline  
\end{tabular}
\caption{\sf \small{Cross section (in $fb$) at the LHC   of
signal and background for the trilepton plus one jet and missing
$p_T$ channel for $R^{-1}$ = 500 GeV.}}
\label{tab:3L}
\end{center}
\end{table}

As for the two-lepton case, here again the $p_T$ of leptons from KK-mode decay
are peaked in the lower range with distributions not extending to high values.
So, it is useful to demand that the $p_T$ of the leptons be confined to a
judiciously chosen window. Further, to remove any $Z$-related background, one
must apply a cut on the invariant mass of all possible pairings of
leptons. Thus, we
are led to the following kinematic cuts:\\

\centerline{
(i)  ${p_T}^{l_2} <$ 25 GeV,  
(ii)  ${p_T}^{l_3} <$ 40 GeV, and 
(iii) $|M_{l_il_j} - M_Z| >$ 10  GeV for $i,j$ = 1,2,3, $i \neq j$.}

The effect of these successive cuts is readily seen from Table \ref{tab:3L}.
For a 10$fb^{-1}$ integrated luminosity with LHC running at $\sqrt{s} = $ 14
(10) TeV, significance which is initially 0.86 (0.43) achieves a respectable
value of 6.39 (3.72) after the kinematic cuts. In Fig.~\ref{lumilim} is shown
(green solid curve) the integrated luminosity necessary for a 5$\sigma$ signal
as a function of $R^{-1}$. It is seen that with 100$fb^{-1}$ data one will
have a reach in $R^{-1}$ of 720 (620) GeV at the 5$\sigma$ level through this
trilepton mode with the LHC running with $\sqrt{s}=$ 14 (10) TeV.

\subsection{Four leptons} 
The signal consisting of four leptons, a jet and missing $p_T$ will arise from
$Q^1 Z^1$ production, followed by the $Q^1$ decaying through a $Z^1$. The
background, in the Standard Model, originates from four $W$-bosons or three
$Z$-bosons, in either case associated with a jet. These processes are expected
to be small. This is borne out by the results presented in Table \ref{tab:4L}.
To enrich the signal with respect to the background all leptons are required
to be isolated ($\Delta R >$ 0.7) from the jet. Further, to eliminate
the background from $Z$-boson decays we require:\\

\centerline{$|M_{l_il_j} - M_Z| >$ 10  GeV for $i,j$ = 1,2,3,4, $i \neq j$.}

\begin{table}[tbh]
\begin{center}
\begin{tabular}{|c|c|c|c|c|}  \hline 
$\sqrt{s}\;\;\;\; \rightarrow$ &\multicolumn{2}{|c|}{14 TeV}&
\multicolumn{2}{|c|}{10 TeV} \\ \hline
Cut used   $ \downarrow$          & Signal       & Background  &
Signal & Background \\  \hline
Basic cuts            &  1.01  & 0.130   & 0.350  & 0.068 \\ \hline 
Lepton isolation    &  0.665  & 0.029   & 0.233  & 0.015 \\ \hline 
$|M_{l_il_j} - M_Z| >$ 10  GeV & 0.573  & 0.004 & 0.206 & 0.002 \\ \hline  
\end{tabular}
\caption{\sf \small{Cross section (in $fb$) at the LHC   of
signal and background for the tetralepton plus one jet and missing
$p_T$ channel for $R^{-1}$ = 500 GeV.}}
\label{tab:4L}
\end{center}
\end{table}

It is seen that the signal and background are both small.  The minimum
integrated luminosity required for a 5$\sigma$ discovery as a function of
$R^{-1}$ is shown in Fig.~\ref{lumilim} (brown dotted curve). With a data set
of 100 $fb^{-1}$ the reach of the four-lepton channel is 600 (500) GeV for the
LHC running with $\sqrt{s} =$ 14 (10) TeV.

\section{Conclusions and Outlook}\label{conclusion}
Universal extra-dimensional models provide a rich spectrum of towers of KK
excitation modes of the SM particles. These KK modes are characterised by the
integers $n$ =1,2,3 \ldots They bear the same quantum numbers as their
zero-mode counterparts but carry higher masses with constant spacing, given by
$n/R$ (upto zero mode masses).  Bearing in mind all the different experimental
constraints on $R$, the lowest (i.e., $n$ = 1) KK excitations can still be
very much within the reach of the LHC. Due to the conservation of KK number at
the tree level vertices which follows from the symmetry of the Lagrangian,
such KK modes are likely to be produced in pairs. Quantum corrections cause
splitting among the different KK states at the same level. The lightest of the
$n$ = 1 states -- the $\gamma^1$ -- is stable and escapes undetected. In this
paper, from the point of view of signal to background optimization (see
Introduction), we have focussed on the production of the $n$ = 1 excitation of
a gauge boson along with an $n$ = 1 excited quark.

The decay of the gauge boson excitations gives rise to leptons and missing
$p_T$ (from the undetected $\gamma^1$) while the quark excitation produces a
jet, missing $p_T$ and possibly leptons. Thus, the signal is a jet, several
leptons and missing $p_T$. The SM background for these final state topologies
is larger than the signal, sometimes overwhelmingly. We have shown that with
judiciously chosen kinematic cuts, including an isolation of the jet from all
leptons, the signal can be enhanced {\em vis-\`{a}-vis} the background, while
retaining enough signal events for a positive verdict with 100 $fb^{-1}$ of
data.  This will be possible so long as $R^{-1}$ does not exceed about
700$-$800 GeV considering that the LHC would have an accumulated luminosity of
(100$-$300) $fb^{-1}$ at $\sqrt{s}$ = 14 TeV.
  
We have classified the cuts in two categories. First, we imposed some basic
cuts to suit LHC observability: ($i$) the leptons are required to satisfy $p_T
>$ 5 GeV, ($ii$) the jet must have a $p_T$ not less than 20 GeV, and ($iii$)
the missing transverse momentum must be more than 25 GeV.  Beyond this, other
kinematic cuts have been appropriately imposed on a case-by-case basis
depending on the number of leptons in the final state. Out of these, two cuts
deserve special mention: ($i$) the requirement of an isolation of the jet from
all leptons is found to be quite useful to remove the top and bottom quark
related backgrounds, and ($ii$) a cut on the lepton-pair invariant mass to
remove on-shell SM $Z$ production backgrounds is also quite effective.  Our
main observation for the specific UED signal cross sections at the LHC with
$\sqrt{s}$ = 14 TeV after the imposition of the kinematic cuts is as follows:
\begin{itemize}
\item
Single jet $+$ ${p}_T\!\!\!\!\!\!/~$ $+$ two leptons:~~ 
Signal: 12.52 $fb$, ~~ Background: 9.18 $fb$,

\item
Single jet $+$ ${p}_T\!\!\!\!\!\!/~$ $+$ three leptons:~~
Signal: 5.00 $fb$, ~~  Background: 1.02 $fb$,

\item
Single jet $+$ ${p}_T\!\!\!\!\!\!/~$ $+$ four leptons:~~
Signal: 0.573 $fb$, ~~ Background: 0.004 $fb$. 
\end{itemize}

The analysis performed here is based on a parton-level simulation and is of an
exploratory nature. For example, it has been assumed that the detectors are of
perfect efficiency, QCD corrections have not been included, and parton
distribution function uncertainties ignored. Our results encourage a detailed
careful analysis with full detector simulation.

It should be observed that the spectrum and the couplings in UED or such
extra-dimensional models are reminiscent of many different non-supersymmetric
scenarios which contain additional gauge bosons and/or vector-like fermions.
A crucial component of UED is the presence of a stable $\gamma^1$ which makes
it different from its peers. The following observation is worth noticing.
Conceptually, UED is closer to the Randall-Sundrum (RS) scenario than
supersymmetry, so one would na\"{i}vely expect similar observational features
between UED and RS than, say, between UED and supersymmetry. First recall the
similarities between UED and RS.  Although, the extra space is warped for RS
and flat for UED, both yield KK modes as a consequence of compactification of
an extra space dimension, and in both cases, the SM particles and their KK
partners share the same spin. This is in contrast to the supersymmetric
extension of the SM (which may be interpreted as a theory with an extra
dimension in fermion coordinates), where the SM particles and their
superpartners carry different spin.  Nonetheless, it turns out that from an
observational point of view, UED is closer to supersymmetry with conserved
$R$-parity (or, for that matter, little Higgs models with conserved
$T$-parity) than RS. This happens primarily because the simplest version of RS
lacks a stable $\gamma^1$ due to the absence of KK parity, while supersymmetry
with conserved $R$-parity (or, little Higgs with conserved $T$-parity) does
contain a stable superparticle (heavy particle). Quite a few LHC simulations
of the RS scenario have been carried out \cite{rslhc}. But due to the absence
of any KK parity in the simplest versions of RS, the KK states, once produced,
decay into the SM particles, and hence the search strategies for RS and UED
would be entirely different. However, as already mentioned in the
Introduction, weak-scale supersymmetry, with a relatively compressed spectrum,
can mimic UED and {\em vice-versa} at LHC.  Distinction between these two new
physics alternatives can only be done by exploiting the spin information
imprinted in angular distributions.  A detailed study of how to differentiate
UED from supersymmetry, following our line of analysis in the context of the
LHC, is beyond the scope of the present work.

\vskip 5pt

{\bf Acknowledgements}: GB and AD are partially supported by project
no. 2007/37/9/BRNS (DAE), India.  AD is also supported in part by CSIR project
no. 03(1085)/07/EMR(II) and the UGC DRS programme (F 530/4/DRS/2009 (SAP 1)).
The work of SKM is partially supported by the Belgian Federal Office for
Scientific, Technical and Cultural Affairs through the Inter-university
Attraction Pole No.  P6/11.  We thank P.~de Aquino for cross-checking some of
our results using the MG and FR-MUED implementation.  SKM is thankful to F.
Maltoni for useful discussions. SKM also thanks Saha Institute of Nuclear
Physics for hospitality at different stages of the work. The research of AR
was supported under the XIth Plan Neutrino Physics and Regional Centre for
Accelerator-based Particle Physics projects at HRI.  The initial computational
work was carried out using the HRI cluster facilities.

\end{document}